\documentclass{aastex631}
\usepackage{amsmath}
\usepackage{booktabs} 
\usepackage{makecell} 
\usepackage{hyperref} 
\usepackage{amsfonts}  
\usepackage{amssymb}   
\usepackage{xcolor}

\begin{document}

\title{Prediction of Large Solar Flares Based on SHARP and HED Magnetic Field Parameters}

\correspondingauthor{Yanfang Zheng}
\email{zyf062856@163.com}
\correspondingauthor{Ting Li}
\email{liting@nao.cas.cn}

\author[0000-0003-0397-4372]{Xuebao Li}
\affiliation{School of  Computer Science, Jiangsu University of Science and Technology, Zhenjiang, 212100, China;}

\author[0009-0000-6353-5473]{Xuefeng Li }
\affiliation{School of Automation, Jiangsu University of Science and Technology, Zhenjiang, 212100, China;}


\author[0000-0003-0229-3989]{Yanfang Zheng}
\affiliation{School of  Computer Science, Jiangsu University of Science and Technology, Zhenjiang, 212100, China;}

\author[0000-0001-6655-1743]{Ting Li}
\affiliation{CAS Key Laboratory of Solar Activity, National Astronomical Observations, Chinese Academy of Sciences,
\\Beijing 100101, China;}
\affiliation{School of Astronomy and Space Science, University of Chinese Academy of Sciences, Beijing 100049, China}
\affiliation{National Space Science Center, Chinese Academy of Sciences, Beijing 100190, China}

\author[0000-0002-3667-3587]{Pengchao Yan}
\affiliation{School of  Computer Science, Jiangsu University of Science and Technology, Zhenjiang, 212100, China;}

\author[0009-0009-9722-5794]{Hongwei Ye}
\affiliation{School of  Computer Science, Jiangsu University of Science and Technology, Zhenjiang, 212100, China;}

\author[0009-0003-0325-1366]{Shunhuang Zhang}
\affiliation{School of  Computer Science, Jiangsu University of Science and Technology, Zhenjiang, 212100, China;}

\author[0009-0004-1408-8055]{Xiaotian Wang}
\affiliation{School of  Automation, Jiangsu University of Science and Technology, Zhenjiang, 212100, China;}

\author[0009-0004-9110-0425]{Yongshang Lv}
\affiliation{School of  Computer Science, Jiangsu University of Science and Technology, Zhenjiang, 212100, China;}

\author[0009-0008-1279-1870]{Xusheng Huang}
\affiliation{School of  Computer Science, Jiangsu University of Science and Technology, Zhenjiang, 212100, China;}


\begin{abstract}
 The existing flare prediction primarily relies on photospheric magnetic field parameters from the entire active region  (AR), such as Space-Weather HMI Activity Region Patches (SHARP) parameters. However, these parameters may not capture the details the AR evolution preceding flares. The magnetic structure within the core area of an AR is essential for predicting large solar flares. This paper utilizes the area of high photospheric free energy density (HED region) as a proxy for the AR core region. We construct two datasets: SHARP and HED datasets. The ARs contained in both datasets are identical. Furthermore, the start and end times for the same AR in both datasets are identical. We develop six models for 24-hour solar flare forecasting, utilizing SHARP and HED datasets. We then compare their categorical and probabilistic forecasting performance. Additionally, we conduct an analysis of parameter importance. The main results are as follows: (1) Among the six solar flare prediction models, the models using HED parameters outperform those using SHARP parameters in both categorical and probabilistic prediction, indicating the important role of the HED region in the flare initiation process. (2) The Transformer flare prediction model stands out significantly in True Skill Statistic (TSS) and Brier Skill Score (BSS), surpassing the other models. (3) In parameter importance analysis, the total photospheric free magnetic energy density ($\mathrm {E_{free}}$) within the HED parameters excels in both categorical and probabilistic forecasting. Similarly, among the SHARP parameters, the R\_VALUE stands out as the most effective parameter for both categorical and probabilistic forecasting.

\end{abstract}

\keywords{Solar activity (1475); Solar flares (1496); Solar active region magnetic fields (1975); Astronomy data analysis(1858)}


\section{Introduction} \label{sec:intro}

Solar flares are powerful explosive events characterized by sudden electromagnetic radiation bursts and the rapid release of high-energy particles
 \citep{fleishman2020decay, sinha2022comparative}. The solar prediction is important because Ultraviolet (UV) and Soft X-ray (SXR) radiation affect the ionosphere and change the radio communication and Low Earth Orbit (LEO) drag \citep{qian2011variability,sadykov2019statistical,monte2014occurrence,nwankwo2018space}. Moreover,
 Solar flares often initiate the chain reaction, which includes coronal mass ejections (CMEs) and solar energetic particle events (SEPs), and therefore their prediction aids the prediction of CMEs and SEPs resulting in a variety of other space weather effects \citep{youssef2012relation,garcia2016prediction,gou2020solar}. Hence, establishing a reliable prediction model for solar flares is crucial for effectively preventing or mitigating the damage caused by solar flares.
\par In recent years, with the accumulation of solar flare observational data and the development of machine learning, some researchers have made progress in predicting solar flares using Space-Weather HMI Activity Region Patches (SHARP; \citealp{bobra2014helioseismic}) data. For instance, some researchers employed multi-layer perceptrons, Support Vector Machines (SVM), and random forest algorithms to predict the occurrence or intensity of solar flares \citep{liu2017predicting,florios2018forecasting,inceoglu2018using,reep2021forecasting}. With the advent of the Recurrent Neural Network (RNN), other researchers utilized the Long Short-Term Memory (LSTM) network, taking into account the time dependency of solar flare prediction \citep{liu2019predicting,sun2019interpreting,wang2020predicting,sun2022predicting}. As image recognition and segmentation technology matured, researchers predicted solar flares by inputting SHARP magnetic maps into a deep convolutional neural network (CNN) \citep{park2018application,li2020predicting,sun2022predicting,deshmukh2022decreasing}. Recently, \cite{liu2019predicting} and \cite{zheng2023multiclass} utilized a combined LSTM-Attention method to enhance the performance of the neural network model for solar flare forecasting. With the continuous iteration and improvement of models, the new Transformer model has emerged as a promising approach for solar flare prediction, leveraging its ability to capture long-range dependencies \citep{kaneda2022flare, abduallah2023operational,pelkum2024forecasting,grim2024solar}.
\par Previous studies have shown that the photospheric non-potential parameters (such as the vertical current, current helicity, shear angle, and so on) are closely related to the occurrence of large solar flares \citep{falconer2002correlation,su2014statistical,toriumi2019flare}. Thus numerous photospheric magnetic field parameters are used to develop flare prediction models \citep{bobra2015solar,liu2017predicting,campi2019feature}.
These widely used parameters are mostly calculated within the entire active region (AR), such as the SHARP parameters. However, they may not capture the finer details preceding solar flare eruptions 
 \citep{barnes2016comparison,toriumi2019flare}. Indeed, the magnetic structure in the core area of an AR is considered a crucial factor in determining whether a large solar flare would occur \citep{schrijver2007characteristic,sun2015great}. For instance, \cite{sun2021improved} indicated that the features extracted within the polarity inversion line (PIL) region which is a sub-region of the AR exhibit greater effectiveness than the SHARP parameters for solar flare prediction. Recently, \cite{li2024survey} used the area of high photospheric free energy density (HED region) as a proxy for the AR core region. They found that the combination of total photospheric free magnetic energy density $E_{free}$  and mean unsigned current helicity $H_{c}$  presents a good ability to distinguish between C-class and M/X-class flaring ARs. 
 In this paper, we establish a database that comprises two datasets: a SHARP dataset constructed using SHARP magnetic field parameters, and an HED dataset focusing on magnetic field parameters of the HED region.
 Additionally, we employ five distinct deep learning algorithms (Transformer, BiLSTM-Attention, BiLSTM, LSTM-Attention, and LSTM) as well as a neural network (NN) model to develop solar flare prediction models for forecasting $\ge$ M class flares within 24 hours. We subsequently utilize the parameters from the SHARP and HED datasets to comparatively evaluate their categorical and probabilistic predictive performance. Finally, we conduct an in-depth analysis to assess the parameter importance of both SHARP and HED parameters, respectively.
\par The rest of this article is organized as follows. The data is described in Section 2, the method is introduced in Section 3. The results are presented in Section 4, and the conclusions and discussions are provided in Section 5.

\section{Data} \label{sec:style}
The database in this paper comprises two datasets: the SHARP dataset and the HED dataset. Specifically, the SHARP dataset incorporates magnetic field parameters derived from SHARP data. Out of the 25 parameters provided in the SHARP data products \citep{bobra2015solar}, 10 parameters have been selected based on prior studies \citep{liu2017predicting,tang2021solar,zheng2023comparative}. These studies make a consensus that these 10 parameters are effective for flare prediction, as depicted in Table \ref{tab1}. The HED dataset comprises magnetic field parameters of the HED region, sourced from \cite{li2024survey}. There are 11 magnetic field parameters that characterize each AR from \cite{li2024survey}. However, some of these parameters exhibit strong correlations among themselves, reflected by exceedingly high Spearman rank-order correlation coefficients. Thus, we narrow down the selection to 6 parameters that exhibit relatively weaker correlations, as detailed in Table  \ref{tab2}. Solar flares are classified as C, M, or X
class according to the peak magnitude of the soft X-ray
flux observed by the Geostationary Operational
Environment Satellite (GOES). This classification of solar flare
is also applied in our work. The solar flare information is obtained from the GOES X-ray flare listings\footnote{\url{https://www.ngdc.noaa.gov/stp/space-weather/solar-data/solar-features/solar-flares/x-rays/goes/xrs/}}.
\par In this work, we extract data from the 24-hour window preceding the solar flare of each AR, capturing 40 data samples per AR at a frequency of 36 minutes, to ensure sufficient variability among the samples of ARs \citep{li2022knowledge,zheng2023multiclass}. The data of \cite{li2024survey} encompasses 323 ARs. However, due to the absence of HED regions in some ARs during this pre-flare period, some ARs may have fewer than the full complement of 40 data samples. We discard those with fewer than 30 samples to maintain the integrity and reliability of our experimental data. Therefore, in this work, we ultimately select 286 ARs, categorizing them into 189 C-class flares and 97 M/X-class flares. The methodology employed for collecting data in the SHARP dataset is aligned with that used for the HED dataset. The SHARP and HED datasets utilized in this paper maintain consistency in the ARs, with start and end times consistent for each AR. Finally, the database for this work comprises both the SHARP and HED datasets, each consisting of 189 C-class flaring ARs, 86 M-class flaring ARs, and 11 X-class flaring ARs.

\par
\begin{figure}[ht!]
\plotone{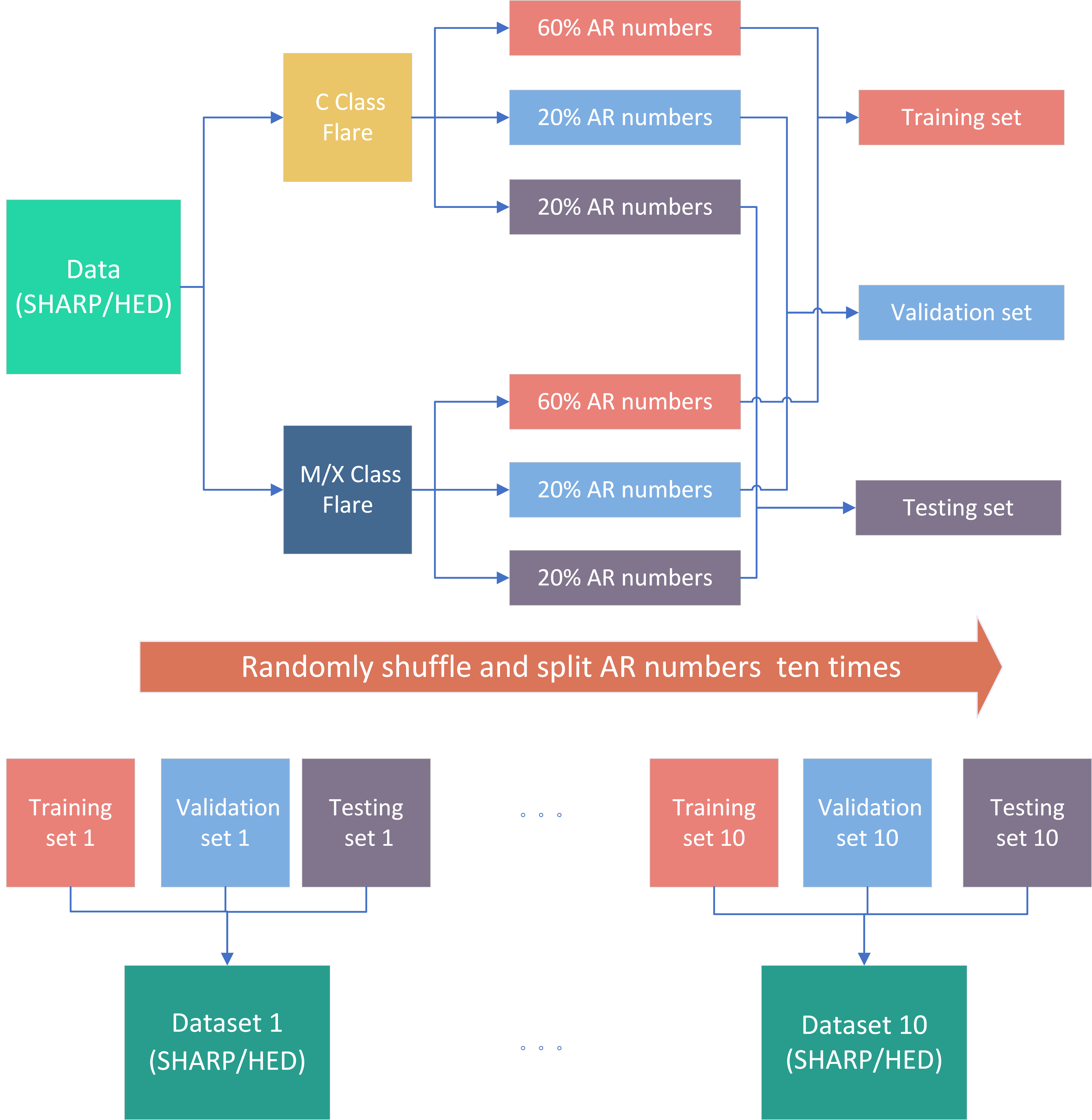}
\caption{The Procedure for Generating 10-Fold Cross-Validation Set Divisions for the SHARP/HED Datasets. 
\label{fig1}}
\end{figure}

 
In this paper, samples of M/X-class $(\ge \mathrm{M})$ flares in an AR are defined as the positive class, whereas C-class samples are defined as the negative class. Since the SHARP and HED parameters have distinct scales and units, they are individually normalized using mean-standard deviation before being input into the flare prediction model. There are three primary normalization strategies: global normalization, local normalization \citep{ahmadzadeh2021train}, and a method that utilizes the mean and standard deviation from the training set to normalize data on the validation and testing sets \citep{jiao2020solar}. According to the research by \cite{ahmadzadeh2021train}, the global normalization method demonstrates superior performance compared to the local normalization method. Therefore, we adopt the global normalization method to normalize the entire dataset. This normalization standardizes each parameter by subtracting its mean and dividing by its standard deviation, ensuring that all parameters have a mean of 0 and a standard deviation of 1, regardless of their original scales. There is no additional transformation (i.e., logarithmic) performed on the parameters.
We employ an AR-based cross-validation (CV) method (\citealp{zheng2019solar, li2020predicting}) to partition the SHARP and HED datasets with the distribution of the training, validation, and testing sets at a ratio of 60\%, 20\%, and 20\%, respectively. This process is repeated ten times using 10 different random seeds, as depicted in Figure \ref{fig1}. This partitioning strategy guarantees that all samples from each individual AR can be arranged only into the training, validation, or testing dataset. This method also allows the model to undergo training, validation, and testing on different CV datasets, ensuring that the model is exposed to a sufficiently diverse range of AR samples. The SHARP and HED datasets are then divided into 10 distinct sets, each comprising training, validation, and testing datasets, respectively. Table \ref{tab3} presents the number of solar magnetogram samples and ARs for a single SHARP/HED dataset, and the other datasets follow the similar distribution pattern. In this paper, the CV sets for the SHARP and HED datasets are precisely aligned to ensure a one-to-one correspondence in their AR distributions. In other words, the distribution of ARs in SHARP Dataset 1 is identical to that in HED Dataset 1, and this consistency extends to all other paired CV sets.

\begin{table}[h]
\centering
\caption{The description and formula of SHARP parameters.}\label{tab1}
\begin{tabular}{lll}
\toprule  
Parameters & \multicolumn{1}{l}{Description}                                        & \multicolumn{1}{l}{Formula}  \\ \midrule 
TOTUSJZ    & Total unsigned vertical current                    & $\mathrm{J_{z_{total} }=\sum \left | J_{Z}  \right |\mathrm{d}A}$        \\
TOTUSJH    & Total unsigned current helicity                    & $\mathrm{H_{C_{total} }\propto \sum \left | B_{Z} \cdot J_{Z}  \right | }$       \\
TOTPOT     & Total photospheric magnetic free energy density    & $\mathrm{\rho  _{tot} \propto \sum (B^{Obs}-B^{Pot}  )^{2} \mathrm{d}A  }$     \\
ABSNJZH    & Absolute value of the net current helicity         & $\mathrm{H_{C_{abs} } \propto \left |\sum  B_{Z} \cdot J_{Z}  \right |  }$     \\
SAVNCPP    & Sum of the modulus of the net current per polarity &$\mathrm{ J_{Z_{sum} } \propto  \left | \sum_{_{}^{} }^{B_{Z}^{+} } J_{Z} \mathrm{d}A  \right | + \left | \sum_{_{}^{} }^{B_{Z}^{-} } J_{Z} \mathrm{d}A  \right |  }$     \\
USFLUX     & Total unsigned flux                                & $\mathrm{\varphi=\sum \left | B_{Z}  \right |\mathrm{d}A }$      \\
AREA\_ACR  & Area of strong field pixels in the active region   & $\mathrm{ Area =\sum Pixels}$        \\
MEANPOT    & Mean photospheric magnetic free energy             & $\mathrm{  \overline{\rho} \propto \frac{1}{N} (B^{Obs} -B^{Pot})^{2}   }$    \\
R\_VALUE   & Sum of flux near polarity inversion line           & $\mathrm{ \varphi = \sum \left| B_{\text{LoS}} \right| \mathrm{d}A \text{ within } R \text{ mask}       }$\\
SHRGT45    & Fraction of area with shear \textgreater{}45°      & $\mathrm{\text{Area with shear } \textgreater \text{ 45°} \text{ / total\_area}}$      \\\bottomrule 
\end{tabular}
\end{table}

\begin{table}[ht]
\centering
\caption{The description and formula of HED parameters.}\label{tab2}
\begin{tabular}{cll}
\toprule 
Parameters & \multicolumn{1}{l}{Description}                & \multicolumn{1}{l}{Formula} \\ \midrule  
$\mathrm{E_{free}} $  & Total photospheric free energy density         & $\mathrm{E_{free} =\sum p_{free} dA    } $                       \\
$\mathrm{\Psi  } $        & Mean shear angle &$\mathrm{ \Psi =\arccos \frac{B_{obs}\cdot B_{pot}  }{\left | B_{obs}B_{pot}  \right | }   } $                       \\
$\mathrm{J_{Z} } $         & Mean unsigned vertical current density         & $\mathrm{J_{Z} =\frac{1}{\mu _{0}N } \sum \left | {\frac{\partial B_{y} }{\partial x} -\frac{\partial B_{x} }{\partial y} }   \right |     } $                       \\
$\mathrm{H_{C}} $         & Mean unsigned current helicity                 &$\mathrm{H_{c} =\frac{\mu _{0} }{N} \sum \left |B_{Z} J_{Z}   \right | } $                           \\
$\mathrm{\alpha } $          & Mean characteristic twist parameter            & $\mathrm{\alpha =\mu _{0} \frac{\sum J_{Z} B_{Z}  }{\sum B_{Z}^{2} }  } $                          \\
$\mathrm{\nabla B_{h} } $         & Mean horizontal gradient of horizontal field   &$\mathrm{ \nabla B_{h} = \frac{1}{N} \sum \sqrt{(\frac{\partial B_{h} }{\partial x} )^{2}+(\frac{\partial B_{h} }{\partial y} )^{2}   }            } $                \\ \bottomrule  
\end{tabular}
\end{table}

\begin{table}[ht]
\centering
\caption{The number of solar magnetogram samples and ARs in one SHARP/HED dataset.}
\label{tab3}
\begin{tabular}{llll}
\toprule  
Data set & \begin{tabular}[c]{@{}l@{}}C class sample/\\ AR numbers\end{tabular} & \begin{tabular}[c]{@{}l@{}}M class sample/\\ AR numbers\end{tabular} & \begin{tabular}[c]{@{}l@{}}X class sample/\\ AR numbers\end{tabular} \\ 
\midrule 
Training set   & \begin{tabular}[c]{@{}l@{}}4520/113\end{tabular}   & \begin{tabular}[c]{@{}l@{}}2120/53\end{tabular}  &\begin{tabular}[c]{@{}l@{}} 240/6 \end{tabular}  \\
Validation set & \begin{tabular}[c]{@{}l@{}}1520/38 \end{tabular}    & \begin{tabular}[c]{@{}l@{}}680/17  \end{tabular} &\begin{tabular}[c]{@{}l@{}} 80/2  \end{tabular}  \\
Testing set    & \begin{tabular}[c]{@{}l@{}}1520/38  \end{tabular}   & \begin{tabular}[c]{@{}l@{}}640/16 \end{tabular}  &\begin{tabular}[c]{@{}l@{}} 120/3 \end{tabular}  \\
Total          & \begin{tabular}[c]{@{}l@{}}7560/189 \end{tabular}   & \begin{tabular}[c]{@{}l@{}}3440/86 \end{tabular} &\begin{tabular}[c]{@{}l@{}} 440/11 \end{tabular} \\
\bottomrule 
\end{tabular}
\end{table}

\section{Method} \label{sec:floats}
In this paper, we design five solar flare prediction models leveraging five currently popular deep learning algorithms: Transformer, BiLSTM-Attention, BiLSTM, LSTM-Attention, and LSTM. The Transformer model consists of an Input layer, a patch layer, a patch encoder layer, Transformer encoder blocks, and an Output layer. Figure \ref{fig2} depicts the architecture of this model. On the other hand, the LSTM series models incorporate an Input layer, an LSTM layer, and an Output layer. Notably, the BiLSTM-Attention and LSTM-Attention models incorporate the attention mechanism, and Figure \ref{fig3} illustrates the architecture of the BiLSTM-Attention model. We use NN model as the baseline model to compare with other deep learning models. The model architecture of the NN model is similar to the output layer of the Transformer and LSTM series models, and uses point-in-time measurements.
\subsection{ Input layer}
In solar flare prediction models, the Input layer remains consistent across all deep learning model architectures. The input sequence for these models is typically denoted as $[\mathrm{X_{t-T+1}, X_{t-T+2}, \ldots, X_{t-1}, X_{t}}]$, where $\mathrm{X_{t}}$ represents the vector input captured at time $\mathrm{t}$, $\mathrm{X_{t-1}}$ represents the vector input at time $\mathrm{t-1}$, and $\mathrm{T}$ designates the time step length which is set to 40 in this paper. Specifically, the Input layer processes $\mathrm{T}$ samples originating from a single AR, with all samples belonging to either the positive or negative class.
\subsection{Transformer layer and LSTM series layer}This paper is inspired by the model structure of Vision Transformer (ViT; \citealp{yuan2021tokens}), which transforms the data output from the Input layer into dimensional representations that are suitable for processing by the Transformer model. This transformation is achieved through the Patch layer and Patch Encoder layer. In this paper, we design a Transformer encoder block. Each Transformer encoder block consists of a Multi-Head Attention (MHA) layer, two Add Layers, a Batch Normalization (BN) layer, and a Multi-Layer Perceptron (MLP) layer. The MHA layer leverages multiple independent self-attention mechanisms simultaneously focusing on different parts of the input sequence, and capturing diverse dependencies \citep{vaswani2017attention}. In this work, the MHA layer is configured with 2 heads, each with 200 dimensions. Add Layer 1 performs residual connections between the input and output of the MHA layer, facilitating better model training and optimization \citep{he2016deep}. The BN layer enhances the stability and convergence speed of neural network training while providing regularization effects that improve the generalization ability of the model \citep{zerveas2021transformer}. The MLP layer further transforms and extracts features, resulting in a more comprehensive representation. In this study, the MLP comprises two Dense layers interspersed with two Dropout layers. Add Layer 2 also performs residual connections, connecting the input of the BN layer to the output of the MLP layer, and feeding the combined result into the next transformer encoder block. We employ a total of 4 transformer encoder blocks.
\par The LSTM layer can be categorized into unidirectional LSTM and bidirectional LSTM. In unidirectional LSTM, the information flows solely in a forward direction, considering only previous data in the sequence to predict the current output. Conversely, in bidirectional LSTM, the flow of information occurs both forward and backward, allowing the model to consider both preceding and following information at each time step. This allows it to produce a more informed output at the current time step. BiLSTM-Attention and LSTM-Attention incorporate an attention mechanism that dynamically assigns varying weights to different inputs of the model. This attention mechanism enables the model to prioritize and focus on the most relevant information, leading to a better understanding and processing of the input data, ultimately resulting in improved model performance.
\subsection{Output layer}
\begin{figure}[ht!]
\plotone{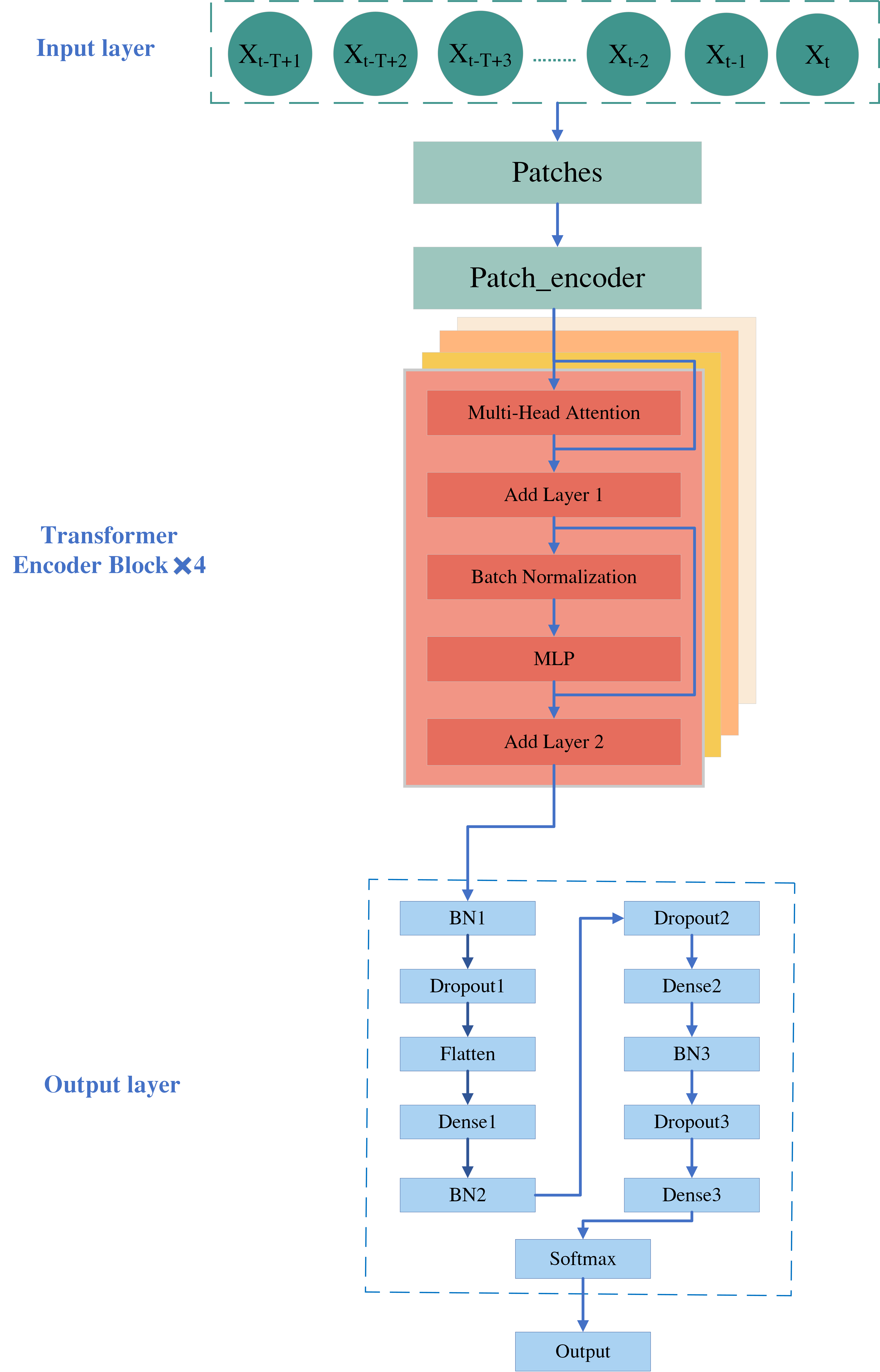}
\caption{The architecture of transformer model. 
\label{fig2}}
\end{figure}

\begin{figure}[ht!]
\plotone{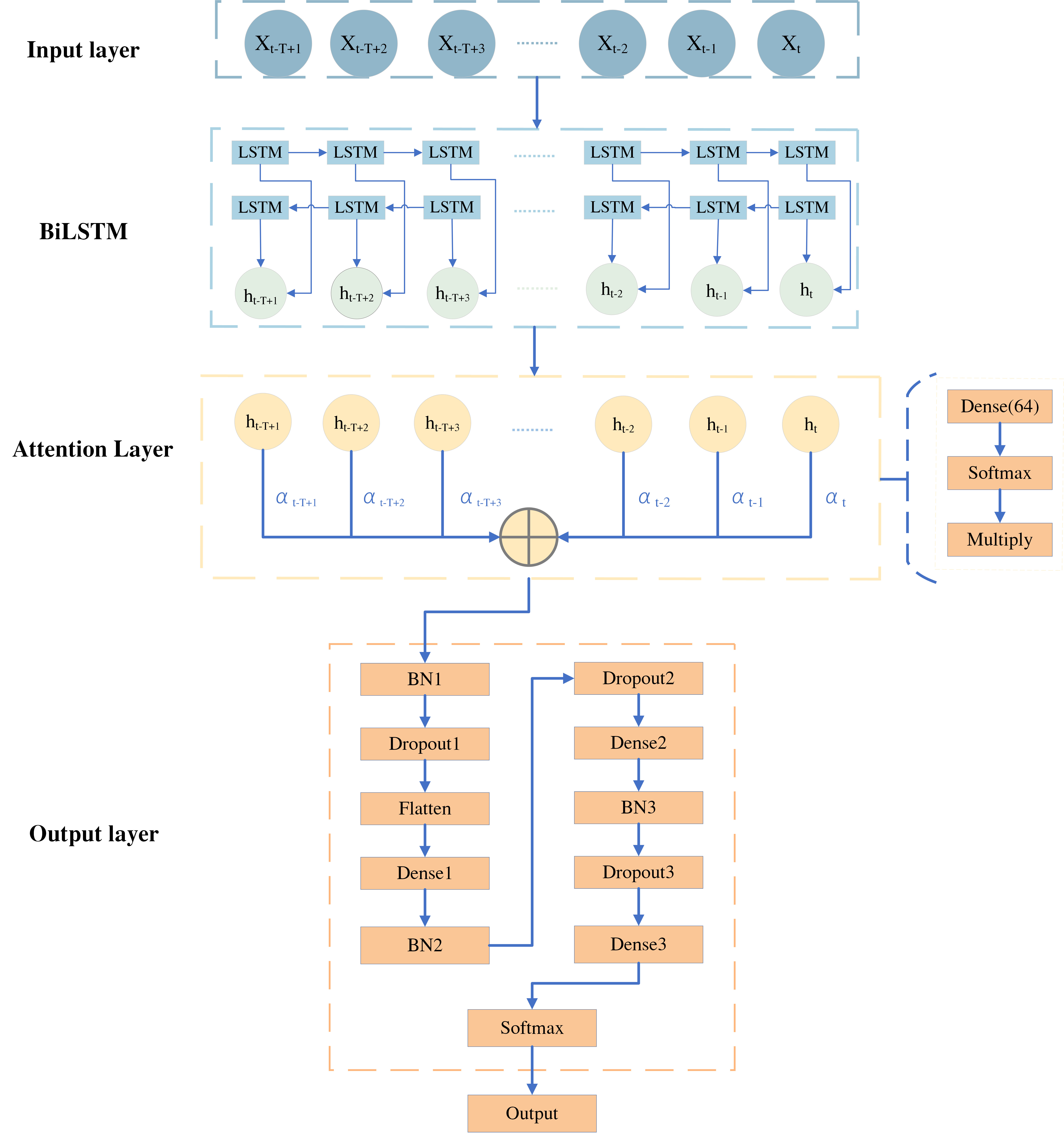}
\caption{The architecture of BiLSTM-Attention model. 
\label{fig3}}
\end{figure}

The output layer consists of three Batch Normalization (BN) layers, three Dense layers,  a Flatten layer, three Dropout functions, and a Softmax activation function. The Dense layer plays a crucial role in neural networks, as it performs linear transformations on the input data, effectively mapping the original feature space into a new one. This transformation facilitates feature extraction and optimizes model training, among other functionalities. The Flatten layer, on the other hand, serves to convert multi-dimensional input data into a flattened one-dimensional representation, ultimately enhancing the utilization of the data. Lastly, the Dropout function randomly deactivates a subset of neurons during training, serving as a regularization technique that prevents overfitting and thereby improves the generalization capability of the model. The Softmax activation function generates two probability values, which sum to 1. Concisely, if the Softmax output is $[{1-\hat{y} _{i} },{\hat{y} _{i} } ]$, then ${1-\hat{y} _{i}}$
represents the probability of $< $ M class, while ${\hat{y} _{i} }$ represents the probability of $\ge \mathrm{ M}$ class. Ultimately, the model categorizes the sample as either a positive class or a negative class, based on the relative magnitudes of these output probabilities. The sequential arrangement of these layers and functions is illustrated in Figures \ref{fig2} and \ref{fig3}. 
\subsection{Model Parameter}
To evaluate their predictive performance, we use 10 distinct datasets from the SHARP dataset and HED dataset, respectively, for training, validation, and testing. Specifically, the training and validation sets are used for model training and validation, while the testing set is exclusively used for model evaluation. In this paper, Hyperopt is used for hyperparameter optimization, which can find the optimal combination of parameters within the range set by the experiment  \citep{feurer2019hyperparameter,yang2020hyperparameter}. The hyperparameter search space and optimal values for categorical prediction using the Transformer model are presented in Table \ref{tab4}. Each model is trained using 150 epochs. The models in this experiment are trained using the Adam optimizer \citep{kingma2014adam}, and the learning rate is obtained through hyperparameter optimization. 
\par To mitigate the issue of class imbalance in the dataset, we utilize a weighted binary cross-entropy (WBCE) loss function during the model training. By introducing the weight coefficient, WBCE gives different weights to positive and negative samples, so that the loss of minority class accounts for a larger proportion of the total loss, thus guiding the model to pay more attention to minority class samples. This function calculates the loss by considering the weighted binary cross-entropy, which measures the difference between the model predictions and the true labels in each iteration. This guides the training process in the correct direction. The loss function is typically expressed as equations \ref{eq1} and \ref{eq2}, where $\omega _{i}$ represents the sample class weights, and $N$ is
the number of training samples in each batch size. $y_{i}$ is the actual label for the $\mathit{i_{th}} $ sample (usually 0 or 1 for binary classification). $\hat{y} _{i}$ represents the probability of $\ge \mathrm{ M}$ class by the model for the $\mathit{i_{th}}   $ sample. $n_{count_{j} }$ is the count of samples for class $j$. $n_{classes}$ is the total number of classes. ${n_{sample}}$ is the number of total training samples.

\begin{equation}\label{eq1}
   \mathrm{ WBCE} = \sum_{i=1}^{N} (\omega _{0}y_{i} \log_{}{\hat{y} _{i} } +\omega_{1} (1-y_{i})\log_{}{(1-\hat{y} _{i})}     ) ,  
\end{equation}

\begin{equation}\label{eq2}
   \omega _{j} =\frac{n_{sample} }{n_{classes}\times n_{count_{j} } }(j=0,1) . 
\end{equation}

\begin{table}[h]
\centering
\caption{The hyperparameter search space and optimal values.}
\label{tab4}
\begin{tabular}{llrr}
\toprule 
\multicolumn{1}{c}{\hspace{-12.5em}\makecell[c]{Parameter\\[-7ex]}\hspace{-4em}}
 & \multicolumn{1}{c}{\hspace{-6.5em}\makecell[c]{Search space\\[-7ex]}\hspace{-3em}} & \multicolumn{2}{c}{\hspace{3.5em}Optimal values} \\  
\cmidrule(lr){3-4}
& & \makecell[c]{SHARP\\parameters} & \makecell[c]{HED\\parameters} \\  
\midrule  
Number of Transformer Blocks & {[}1, 2, 3, 4{]}                          & 4                 & 4              \\
Number of Heads in MHA       & {[}1, 2, 3, 4{]}                          & 2                 & 2              \\
Learning Rate                & {[}0.0001 $\sim$ 0.01{]}                 & 0.0020            & 0.0001         \\
MLP Dropout Rate             & {[}0.1 $\sim$ 0.5{]}                     & 0.3200            & 0.3000         \\
Dropout Rate 1               & {[}0.1 $\sim$ 0.5{]}                     & 0.2000            & 0.3990         \\
Dropout Rate 2               & {[}0.1 $\sim$ 0.5{]}                     & 0.1000            & 0.1001         \\
Dropout Rate 3               & {[}0.1 $\sim$ 0.5{]} & 0.1000            & 0.1000         \\
Batch Size                   & {[}5, 10, 15, 20{]}     & 10                & 10             \\ 
Epochs                   & {[}100, 120, 150, 200{]}     & 150                & 150             \\ 
\bottomrule
\end{tabular}
\end{table}

\section{Results} \label{subsec:tables}
In this paper, we take solar flare forecasting as a binary classification task and evaluate the model performance in categorical and probabilistic forecasting. We leverage a confusion matrix to assess the categorical performance of the model. Specifically, samples accurately classified as the positive class are defined as true positives (TP). Conversely, samples accurately classified as the negative class are defined as true negatives (TN). On the flip side, samples erroneously classified as the positive class are defined as false positives (FP). Similarly, samples erroneously classified as the negative class are defined as false negatives (FN). These four quantities constitute the confusion matrix, which serves as the basis for calculating metrics such as Precision, Recall, Accuracy, False Alarm Ratio (FAR), Heidke Skill Score (HSS), and True Skill Statistics (TSS), as detailed in Table \ref{tab5}. 
We use Random chance forecast and Negative forecast, as reference points to compare the performance of the categorical prediction. In the Random chance forecast, TP=${\frac{P}{P+N} \times P}$, where $P$ represents the number of positive samples and $N$ represents the number of negative samples in the testing set. FN=$P$-TP, TN=${\frac{N}{P+N} \times N}$, and FP=${N}$-TN. In the Negative forecast, TP and FP are both equal to 0, FN is equal to $P$, and TN is equal to $N$. By substituting these values for the testing set into the confusion matrix, the corresponding categorical statistic score is obtained.
Given that most of these metrics, except for TSS, are susceptible to imbalances in the experimental sample data \citep{bobra2015solar,bloomfield2012toward}, we designate TSS as the primary evaluation metric in the categorical performance.
\begin{table}[h]
\centering
\caption{Categorical statistic score computed from the confusion matrix.}
\label{tab5}
\begin{tabular}{ccrccc}
\toprule 
Metrics   & Definition & \multicolumn{1}{c@{}}{ Scale}  & \multicolumn{1}{c@{}}{Perfect} & \multicolumn{1}{c@{}}{Random chance forecast}& \multicolumn{1}{c@{}}{Negative forecast}\\\midrule 
Precision & $\mathrm{\frac{TP}{TP+FP}}$   &  0  $\sim$   1     & 1  & 0.333  & /   \\
Recall    & $\mathrm{\frac{TP}{TP+FN}}$            &0 $\sim$   1       & 1   & 0.333   &0 \\
Accuracy  & $\mathrm{\frac{TP+TN}{TP+FP+TN+FN}}$           &0 $\sim $  1       & 1   & 0.556  & 0.667  \\
FAR       & $\mathrm{\frac{FP}{TP+FP}}$             & 0 $\sim$   1      & 0     & 0.667  & / \\
HSS       & $\mathrm{\frac{2[(TP\times TN )-(FN\times FP  )]}{(TP+FN)(FN+TN)+(TP+FP)(FP+TN)}}$            & $ -\infty$ $\sim$   1   & 1  & 0    & 0 \\
TSS       &  $\mathrm{\frac{TP}{TP+FN} -\frac{FP}{FP+TN}} $          &-1 $\sim$   1       & 1 & 0 & 0\\
\bottomrule  
\end{tabular}
\end{table}

\par In addition, we utilize the Brier Score (BS; \citealp{brier1950verification}) and the Brier Skill Score (BSS; \citealp{wilks2010sampling}) as quantitative metrics to evaluate the probabilistic forecasting performance of our models. The formulations for BS and BSS are presented in equations \ref{eq3} and \ref{eq4}, respectively. Specifically, $N$ denotes the total number of data samples in the testing dataset. $y_{i}$ represents the target label, while $\hat{y} _{i}$ is the predicted probability of the $\ge$ M class for the $\mathit{i_{th}}$ testing sample. $\bar{y} =\frac{1}{N   }  {\textstyle \sum_{i=1}^{N}y_{i} } $ indicates the average target label value for the $\ge$ M class across all testing samples. BS ranges from 0 to 1, with a score of 0 representing the best probabilistic forecasting performance. Conversely, BSS ranges from $-\infty $ to 1, with a value of 1 indicating the optimal probabilistic forecasting performance.
\begin{equation}
\label{eq3}
BS=\frac{1}{N}{\textstyle \sum_{i=1}^{N}}(y_{i}-\hat{y} _{i}) ^{2} \text{,}  
\end{equation}

\begin{equation}
\label{eq4}
BSS= 1 - \frac{BS}{\frac{1}{N} \sum_{i=1}^{N}(y_{i}-\bar{y})^{2}} .
\end{equation}
\subsection{Model performance evaluation}
To ensure that the model does not underfit or overfit the data, we take the Transformer model using HED parameters as an example. In the process of model training for categorical prediction, we monitor the TSS on the validation set for each epoch and save the model that corresponds to the optimal TSS value achieved on the validation set. Similarly, during the process of model training for probability prediction, we track the BSS on the validation set per epoch, and retain the model corresponding to the best BSS value obtained.
\par
\begin{figure}[ht!]
\plotone{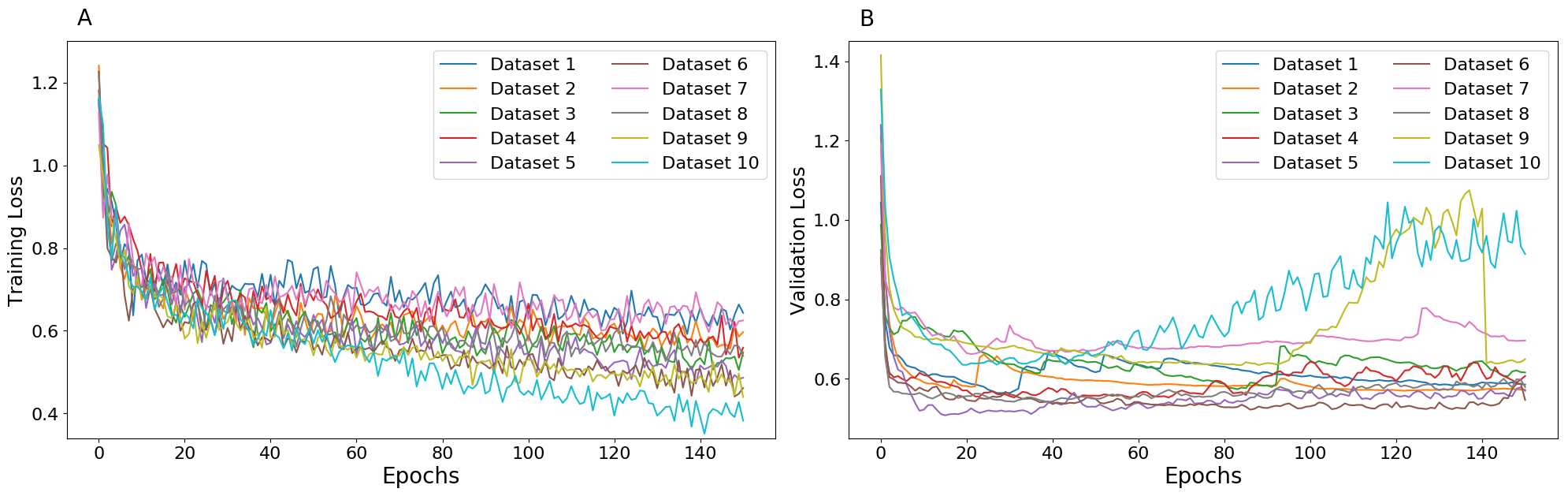}
\caption{The figure depicts the training loss curves and validation loss curves of a Transformer model using HED parameters in $\ge $ M class flare prediction. Panel (A) signifies the training loss curves, while panel (B) denotes the validation loss curves. The 10 different colored curves represent the changes of training and validation loss with epochs for the model trained and validated by 10 separate datasets.}
\label{fig4}
\end{figure}

 As the model architectures and hyperparameters utilized for both categorical forecasting and probability forecasting are identical, the only distinction is the specific epochs at which the optimal models are saved. Therefore, the training loss curves for both the categorical forecasting model and the probability forecasting model are consistent, and the validation loss curves between them are also the same, as shown in Figure \ref{fig4}. Although the training loss curves and the validation loss curves exhibit some fluctuations, they are finally convergent, thus indicating no underfitting situation. Notably, despite the observed overfitting in the validation loss curves 9 and 10 in Figure \ref{fig4}, the model we ultimately saved does not exhibit overfitting. In particular, we take the training loss curve 9 as an example. During the training process for categorical forecasting, although the validation loss curve indicates overfitting between epochs 95 and 145, the model is saved at epoch 150 corresponding to the optimal TSS value achieved on the validation set. In this epoch, the validation loss of the model is still very low, indicating that overfitting has disappeared. Similarly, in the training for probability forecasting, the model that holds the optimal BSS values at epoch 84 does not overfit, essentially equivalent to an earlier termination of the training process. Hence, by monitoring the optimal TSS and BSS values on the validation set, we ensure that the models saved from this experiment exhibit neither underfitting nor overfitting.
\par
\begin{figure}[ht!]
\plotone{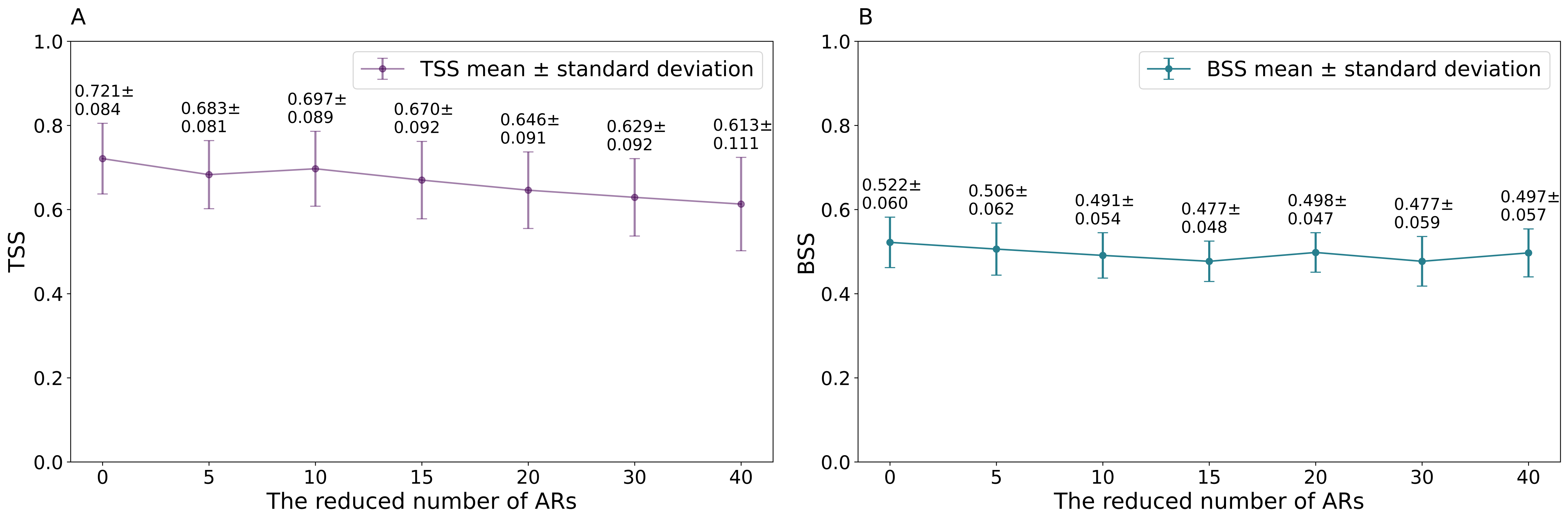}
\caption{The line chart illustrates the mean and standard deviation of TSS and BSS during the reduction of AR numbers in predicting $\ge $ M-class flares. Panel (A) represents the mean and standard deviation of TSS as the number of AR gradually decreases. Panel (B) shows the mean and standard deviation of BSS as the number of AR gradually decreases. The X-axis coordinates represent the reduced number of ARs. }
\label{fig5}
\end{figure}

 To assess the influence of the quantity of unique flares on model training, we conduct an experiment leveraging the Transformer model with HED parameters. Specifically, we randomly reduce the number of ARs within the training set while maintaining the number of ARs in both the validation and testing sets unchanged, and then retrain the model. By observing the changes of TSS in categorical prediction and BSS in probability prediction, we can verify whether the quantity of ARs in the training set is sufficient for training the Transformer model. Figure \ref{fig5} illustrates the mean and standard deviation of TSS and BSS during the reduction of AR numbers. As depicted in Figure \ref{fig5}(A), a slight fluctuation in the TSS score is observed as the number of ARs in the training set decreases in the range of 0 to 15. However, as the number of ARs in the training set reduces by more than 15, there is a slight decline in TSS score. As shown in Figure \ref{fig5}(B), as the number of ARs in the training set decreases, the variation in BSS remains minimal. These results indicate that the number of ARs in the training dataset is sufficient to support the training of the Transformer model. Subsequently, we evaluate the performance of six solar flare prediction models in predicting solar flares of $\ge$ M class on ten testing datasets to determine whether SHARP parameters or HED parameters are more effective in distinguishing between C-class and M/X-class flares.

Table \ref{tab6} presents the categorical results of six models for binary-class predictions within a 24-hour window. The mean TSS score of the six solar flare prediction models using SHARP parameters ranges from 0.376 to 0.559, with the Transformer model achieving the best performance, surpassing other models with a score of 0.559 $\pm $ 0.087.  On the other hand, when using HED parameters, the mean TSS score of the six solar flare prediction models ranges from 0.643 to 0.721.  The Transformer model again outperforms the others, scoring 0.721 $\pm $ 0.084. The BiLSTM model performs the worst with a score of 0.643 $\pm $ 0.074, which is still much better than the best-performing Transformer model using SHARP parameters. We observe that the Transformer model performed the best in both SHARP parameters and HED parameters. In addition, in each of the six models, the categorical forecasting performance using HED parameters is superior to that using SHARP parameters.
\par Table \ref{tab7} shows the probabilistic forecasting results of six models. Among the six solar flare prediction models using SHARP parameters, the Transformer model exhibits the best performance in probabilistic forecasting, achieving a BSS score of 0.304 $\pm $ 0.092, while the other models performed comparably with a mean BSS score ranging from 0.080 to 0.161. Meanwhile, the Transformer model stands out in the six solar flare prediction models using HED parameters, achieving a BSS score of 0.522 $\pm $ 0.060, significantly outperforming the other models. The other models also performed comparably, with a mean BSS score ranging from 0.282 to 0.446, which is still almost superior to the probabilistic results using SHARP parameters. It is observed that, when using the same type of parameters, the Transformer model consistently showed superior performance in probabilistic forecasting compared to other models. Furthermore, for the same model, the probabilistic forecasting results using HED parameters are significantly better than those using SHARP parameters. 

\par Overall, the advanced deep learning models, which enable the extraction of intricate relationships among AR samples, can obtain better categorical and probability forecast performance than the NN model, yielding superior forecasting results for solar flares. Moreover, the Transformer model consistently outperforms other models in both categorical and probabilistic predictions, regardless of whether SHARP parameters or HED parameters are used.
We infer that this may be because the LSTM model processes time series through recursion, but its ability to handle long-range dependencies is limited by its recursive structure. Although the bidirectional LSTM model is capable of capturing contextual relationships through both forward and backward information, it is still limited to capturing local contextual relationships \citep{schmidhuber1997long,sherstinsky2020fundamentals}. However, the Transformer model, endowed with its unique self-attention mechanism, allows direct access to the relationship between any two positions within the sequence, thereby enabling it to learn comprehensive global contextual information. As a result, the Transformer model excels at capturing not only long-range dependencies but also intricate relationships and patterns that exist between different positions within a sequence \citep{vaswani2017attention}. Therefore, the Transformer-type architecture exhibits exceptional performance. Furthermore, the performance of all models utilizing HED parameters surpasses that of models using SHARP parameters in both categorical and probabilistic prediction results.
\subsection{Parameter importance analysis}
\begin{figure}[h!]
\plotone{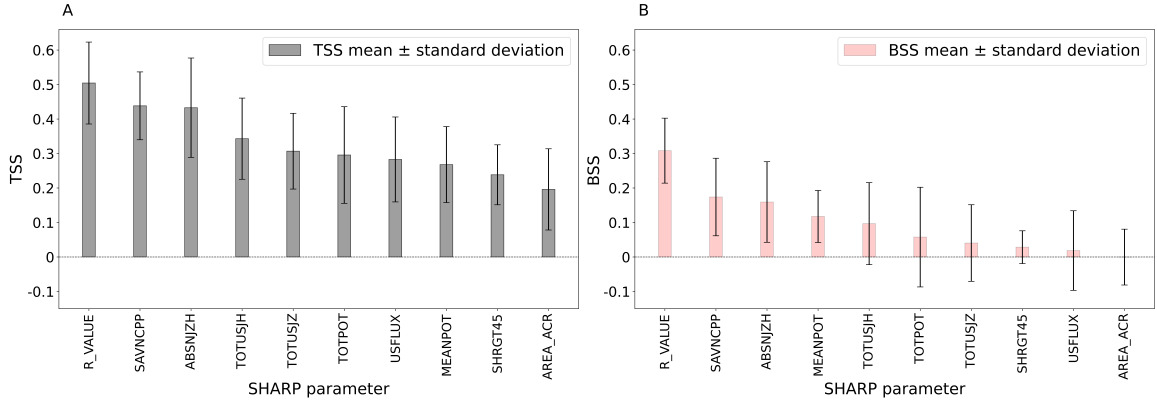}
\caption{The histogram depicts the mean and standard deviation of TSS and BSS for each individual SHARP parameter in $\ge $ M class flare prediction. Specifically, Panel (A) signifies the mean and standard deviation of TSS, while panel (B) denotes the mean and standard deviation of BSS, both of which are sorted in descending order. The ’bar’ denotes the standard deviation.
\label{fig6}}
\end{figure}

\begin{figure}[h!]
\plotone{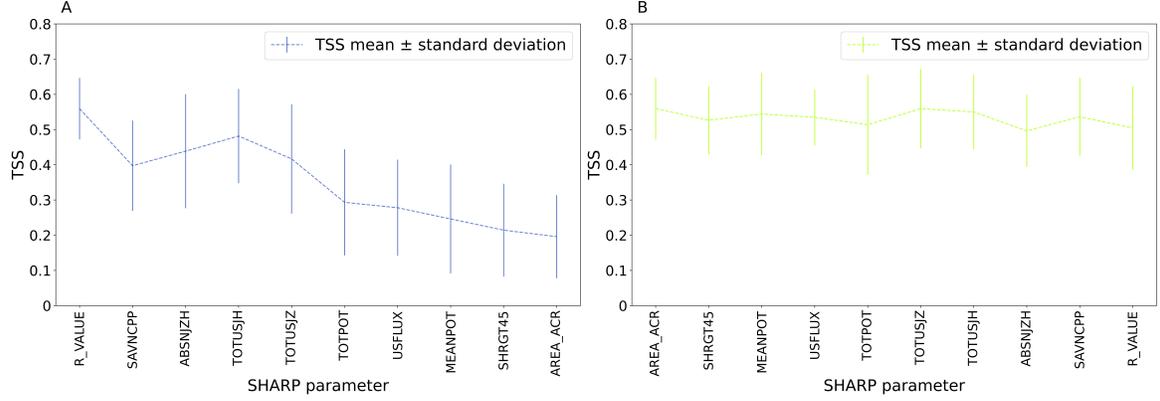}
\caption{The line chart illustrates the mean and standard deviation of TSS during the sequential reduction of SHARP parameters in predicting $\geq$ M-class flares. Panel (A) represents the mean and standard deviation of TSS as the most important SHARP parameters are sequentially removed, starting from the most important. Panel (B) shows the mean and standard deviation of TSS as the least important SHARP parameters are sequentially removed, starting from the least important. The X-axis coordinates represent the combination of a parameter with all parameters to its right.
\label{fig7}}
\end{figure}

In this paper, given that the SHARP parameters and HED parameters performed optimally in the Transformer model, we utilize the Transformer model and apply Recursive Feature Elimination (RFE; \citealp{abhale2023enhancing}) to conduct separate parameter importance analyses for SHARP and HED parameters, respectively. The study by \cite{bobra2016predicting} found that combining certain lower-ranked parameters could enhance model performance. Therefore, we investigate the effects of various parameter combinations on model performance by employing the RFE method. The steps of RFE are as follows. (1) In parameter importance analysis, we solely focus on the TSS score of categorical prediction performance and the BSS score of probabilistic prediction performance for each parameter. Specifically, we train, validate, and test the Transformer model on univariate datasets to evaluate categorical and probabilistic forecast performance, respectively. Therefore, the TSS and BSS scores for the univariate prediction are obtained. Then, we rank the TSS score for each parameter in descending order, and repeat this process for BSS score as well. The TSS and BSS scores for the individual SHARP and HED parameters are ranked in descending order, as shown in Figures \ref{fig6} and \ref{fig9}, respectively. (2) We sort individual parameters according to their TSS and BSS scores in descending and ascending order, respectively. The descending order is arranged from the highest score (most important) to the lowest score (least important), whereas the ascending order follows the opposite direction. Subsequently, we change the dimension of input parameters sequentially, following descending and ascending order. Taking the categorical forecast performance of the SHARP parameters as an example, in the process of sequentially removing the most important parameters, as shown in Figure \ref{fig7}(A), the R\_VALUE on the X-axis represents the initial set of 10 SHARP parameters, SAVNCPP signifies the set with R\_VALUE excluded, ABSNJZH represents the set with both R\_VALUE and SAVNCPP removed, and so forth. Conversely, in the process of sequentially removing the least important parameters, as shown in Figure \ref{fig7}(B), AREA\_ACR on the X-axis represents the original set of 10 SHARP parameters, SHRGT45 denotes the set with AREA\_ACR omitted, MEANPOT indicates the set further excluding SHRGT45, and so on. (3) We train, validate, and test the model under different combinations of SHARP and HED parameters. As elucidated in the study conducted by \cite{liu2020SDOHMI}, an exhaustive search for the optimal parameter subset among $n$ parameters would require evaluating all $2^{n}- 1$ combinations. In light of the extensive parameter space in our study, which comprises 10 SHARP parameters resulting in 1023 combinations and 6 HED parameters yielding 63 combinations, an exhaustive search for the optimal parameter subset becomes computationally intractable. Therefore, we do not consider all possible combinations but rather employ a more targeted RFE approach.

\par
\begin{figure}[ht!]
\plotone{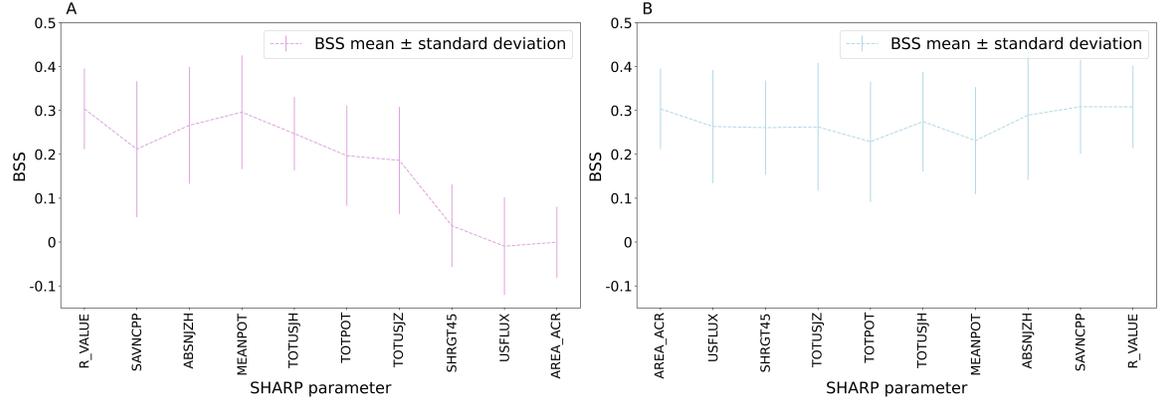}
\caption{The line chart illustrates the mean and standard deviation of BSS during the sequential reduction of SHARP parameters in predicting $\geq$ M-class flares. Panel (A) represents the mean and standard deviation of BSS as the most important SHARP parameters are sequentially removed, starting from the most important. Panel (B) shows the mean and standard deviation of BSS as the least important SHARP parameters are sequentially removed, starting from the least important. The X-axis coordinates represent the combination of a parameter with all parameters to its right.
\label{fig8}}
\end{figure}

\begin{figure}[ht!]
\plotone{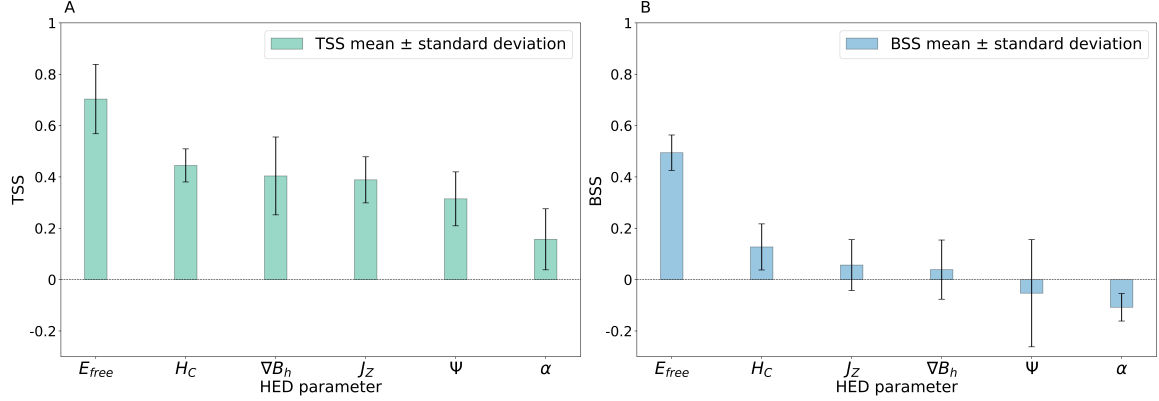}
\caption{The histogram depicts the mean and standard deviation of TSS and BSS for each individual HED parameter in $\ge $ M class flare prediction. Specifically, panel (A) signifies the mean and standard deviation of TSS, while panel (B) denotes the mean and standard deviation of BSS, both of which are sorted in descending order. The ’bar’ denotes the standard deviation.
\label{fig9}}
\end{figure}

 Figure \ref{fig6} shows a histogram that depicts the mean and standard deviation of TSS and BSS for the single SHARP parameter in  $\ge $ M class flare prediction. As depicted in Figure \ref{fig6}, when ranking the parameter importance of individual parameters within the SHARP parameters, the top three most significant parameters for both TSS and BSS scores are consistently R\_VALUE, SAVNCPP, and ABSNJZH. Figure \ref{fig7} shows a line chart displaying the mean and standard deviation of TSS as each SHARP parameter is sequentially reduced in the flare prediction of $\ge $ M class flares. As shown in Figure \ref{fig7}(A), with the removal of the top-ranked parameter, R\_VALUE, the TSS score experiences a notable decline. Meanwhile, as illustrated in Figure \ref{fig7}(B), after removing the six least significant parameters, the categorical forecasting performance of the model with the remaining parameter combination remains high. Figure \ref{fig8} shows a line chart displaying the mean and standard deviation of BSS as each SHARP parameter is sequentially reduced in the flare prediction of $\ge $ M class flares. As observed in Figure \ref{fig8}(A), the BSS score experiences a significant drop when parameters are eliminated down to the less important sets, such as SHRGT45, USFLUX, and AREA\_ACR. However, as depicted in Figure \ref{fig8}(B), when removing the least significant parameters from the SHARP parameters sequentially, the BSS curve exhibits minimal fluctuation. This indicates that the categorical forecasting performance and the probabilistic forecasting performance of the parameter R\_VALUE are both excellent and play a pivotal role in predicting $\ge $ M class flares, which aligns with previous findings by \cite{schrijver2007characteristic}, \cite{welsch2009relationship}, and \cite{liu2017predicting}. 
\par
\begin{figure}[ht!]
\plotone{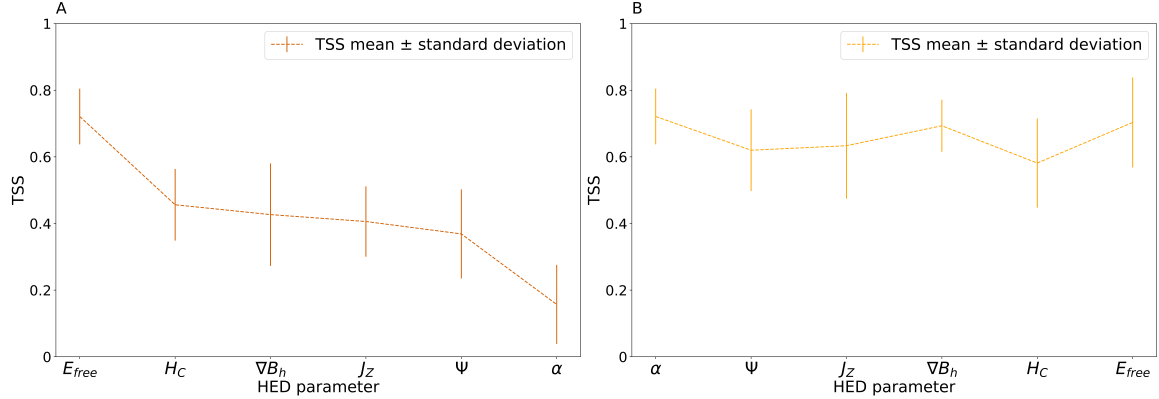}
\caption{The line chart illustrates the mean and standard deviation of TSS during the sequential reduction of HED parameters in predicting $\geq$ M-class flares. Panel (A) represents the mean and standard deviation of TSS as the most important HED parameters are sequentially removed, starting from the most important. Panel (B) shows the mean and standard deviation of TSS as the least important HED parameters are sequentially removed, starting from the least important. The X-axis coordinates represent the combination of a parameter with all parameters to its right.}
\label{fig10}
\end{figure}

\begin{figure}[ht!]
\plotone{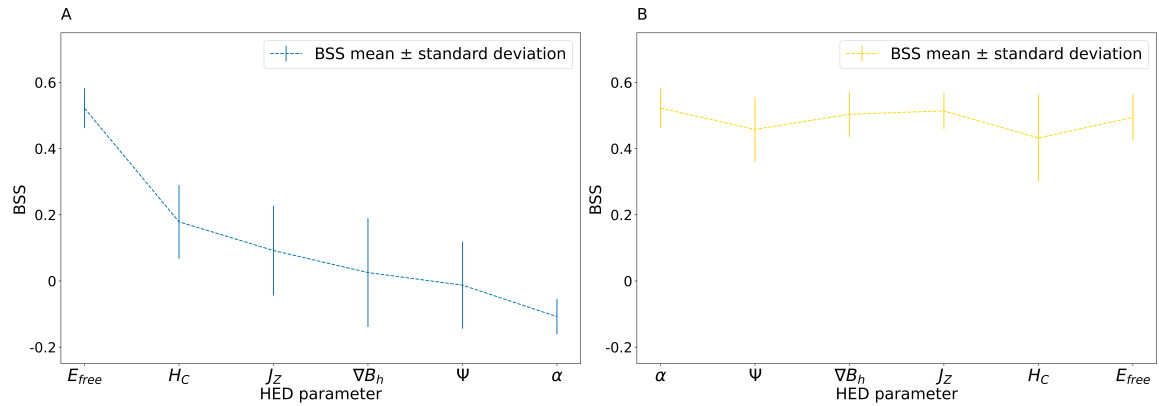}
\caption{The line chart illustrates the mean and standard deviation of BSS during the sequential reduction of HED parameters in predicting $\geq$ M-class flares. Panel (A) represents the mean and standard deviation of BSS as the most important HED parameters are sequentially removed, starting from the most important. Panel (B) shows the mean and standard deviation of BSS as the least important HED parameters are sequentially removed, starting from the least important. The X-axis coordinates represent the combination of a parameter with all parameters to its right.
\label{fig11}}
\end{figure}

 Figure \ref{fig9} shows a histogram that depicts the mean and standard deviation of TSS and BSS for the single HED parameter in  $\ge $ M class flare prediction. As shown in Figure \ref{fig9}, in the ranking of parameter importance among individual parameters in HED parameters, $\mathrm{E_{free}}$ and $\mathrm{H_{C}}$ emerge as the top two parameters based on their TSS and BSS scores. Figure \ref{fig10} shows a line chart displaying the mean and standard deviation of TSS as each HED parameter is sequentially reduced in the prediction of $\ge $ M class flares. In Figure \ref{fig10}(A), a noticeable decline in the TSS score is observed with the removal of $\mathrm{E_{free}}$ and $\mathrm{\Psi}$ from the parameters, respectively. Conversely, as depicted in Figure \ref{fig10}(B), the TSS curve remains relatively stable when the least significant HED parameters are eliminated sequentially. Figure \ref{fig11} shows a line chart displaying the mean and standard deviation of BSS as each HED parameter is sequentially reduced in the prediction of $\ge $ M class flares. Similarly, in Figure \ref{fig11}(A), the BSS score undergoes a marked drop with the exclusion of $\mathrm{E_{free}}$. However, in Figure \ref{fig11}(B), the BSS curve remains largely unaffected when the least important HED parameters are removed sequentially. This indicates that $\mathrm{E_{free}}$ plays a crucial role in $\ge $ M class flare prediction and performs well in both categorical forecasting and probabilistic forecasting, aligning with the conclusion drawn by \cite{li2024survey}.
\par In general, our analysis shows that R\_VALUE is the top performer among SHARP parameters for flare prediction of $\ge $ M class flares, while $\mathrm{E_{free}}$ significantly outperforms the other HED parameters. Furthermore, we notice a tendency that the Transformer model, when using parameters with higher TSS scores in categorical prediction, also tends to achieve higher BSS scores in probabilistic prediction.

\begin{table}[h]
\centering
\caption{The categorical results of the six models for binary-class predictions within 24 h. The boldface represents the best value in each column. The number before the $\pm$ represents the mean, while the number after the $\pm$ represents the standard deviation.}
\label{tab6}
\begin{tabular}{cccc}
\toprule 
\multicolumn{1}{c}{\hspace{-3em}\makecell[c]{Metric\\[-6ex]}\hspace{1em}}  & \multicolumn{1}{c}{\hspace{-1em}\makecell[c]{Model\\[-6ex]}\hspace{1em}} & \multicolumn{2}{c}{\makecell[c]{Results}} \\  
\cmidrule(lr){3-4}  
& & \makecell[c]{SHARP\\parameters} & \makecell[c]{HED\\parameters} \\  
\midrule  

TSS       & \begin{tabular}[c]{@{}c@{}}Transformer\\ BiLSTM-Attention\\ BiLSTM\\ LSTM-Attention\\ LSTM\\ NN\end{tabular} & \begin{tabular}[c]{@{}c@{}}\textbf{0.559 ± 0.087}\\ 0.541 ± 0.127\\ 0.503 ± 0.135\\ 0.510 ± 0.135\\ 0.521 ± 0.118\\ 0.376 ± 0.093\end{tabular} & \begin{tabular}[c]{@{}c@{}}\textbf{0.721 ± 0.084}\\ 0.692 ± 0.075\\ 0.643 ± 0.074\\ 0.705 ± 0.076\\ 0.686 ± 0.065\\ 0.666 ± 0.063\end{tabular} \\\midrule
HSS       & \begin{tabular}[c]{@{}c@{}}Transformer\\ BiLSTM-Attention\\ BiLSTM\\ LSTM-Attention\\ LSTM\\ NN\end{tabular} & \begin{tabular}[c]{@{}c@{}}0.501 ± 0.091\\ \textbf{0.504 ± 0.121}\\ 0.450 ± 0.125\\ 0.457 ± 0.117\\ 0.468 ± 0.107\\ 0.380 ± 0.089\end{tabular} & \begin{tabular}[c]{@{}c@{}}\textbf{0.709 ± 0.084}\\ 0.655 ± 0.081\\ 0.591 ± 0.091\\ 0.680 ± 0.087\\ 0.663 ± 0.058\\ 0.662 ± 0.061\end{tabular} \\\midrule
FAR       & \begin{tabular}[c]{@{}c@{}}Transformer\\ BiLSTM-Attention\\ BiLSTM\\ LSTM-Attention\\ LSTM\\ NN\end{tabular} & \begin{tabular}[c]{@{}c@{}}0.592 ± 0.075\\\textbf{ 0.379 ± 0.086}\\ 0.437 ± 0.074\\ 0.429 ± 0.064\\ 0.425 ± 0.063\\ 0.598 ± 0.058\end{tabular} & \begin{tabular}[c]{@{}c@{}}\textbf{0.210 ± 0.074}\\ 0.282 ± 0.076\\ 0.340 ± 0.094\\ 0.248 ± 0.096\\ 0.261 ± 0.051\\ 0.231 ± 0.056\end{tabular} \\\midrule
Accuracy  & \begin{tabular}[c]{@{}c@{}}Transformer\\ BiLSTM-Attention\\ BiLSTM\\ LSTM-Attention\\ LSTM\\ NN\end{tabular} & \begin{tabular}[c]{@{}c@{}}0.751 ± 0.054\\ \textbf{0.763 ± 0.064}\\ 0.728 ± 0.065\\ 0.735 ± 0.053\\ 0.739 ± 0.054\\ 0.730 ± 0.040\end{tabular} & \begin{tabular}[c]{@{}c@{}}\textbf{0.869 ± 0.038}\\ 0.838 ± 0.047\\ 0.801 ± 0.054\\ 0.852 ± 0.045\\ 0.846 ± 0.027\\ 0.850 ± 0.028\end{tabular} \\\midrule
Precision & \begin{tabular}[c]{@{}c@{}}Transformer\\ BiLSTM-Attention\\ BiLSTM\\ LSTM-Attention\\ LSTM\\ NN\end{tabular} & \begin{tabular}[c]{@{}c@{}}0.408 ± 0.076\\ \textbf{0.621 ± 0.086}\\ 0.563 ± 0.074\\ 0.571 ± 0.064\\ 0.575 ± 0.063\\ 0.402 ± 0.058\end{tabular} & \begin{tabular}[c]{@{}c@{}}\textbf{0.790 ± 0.074}\\ 0.718 ± 0.076\\ 0.659 ± 0.094\\ 0.752 ± 0.096\\ 0.739 ± 0.051\\ 0.769 ± 0.056\end{tabular} \\\midrule
Recall    & \begin{tabular}[c]{@{}c@{}}Transformer\\ BiLSTM-Attention\\ BiLSTM\\ LSTM-Attention\\ LSTM\\NN\end{tabular} & \begin{tabular}[c]{@{}c@{}}\textbf{0.865 ± 0.089}\\ 0.791 ± 0.135\\ 0.821 ± 0.105\\ 0.813 ± 0.165\\ 0.827 ± 0.121\\ 0.568 ± 0.105\end{tabular} & \begin{tabular}[c]{@{}c@{}}0.834 ± 0.080\\ 0.871 ± 0.077\\ \textbf{0.882 ± 0.071}\\ 0.855 ± 0.075\\ 0.837 ± 0.074\\ 0.784 ± 0.060\end{tabular} \\ \bottomrule
\end{tabular}
\end{table}

\begin{table}[h]
\centering
\caption{The probabilistic results of the six models for binary-class predictions within 24 h. The boldface represents the best value in each column. The number before the $\pm$ represents the mean, while the number after the $\pm$ represents the standard deviation.}
\label{tab7}
\begin{tabular}{cccc}
\toprule 
\multicolumn{1}{c}{\hspace{-3em}\makecell[c]{Metric\\[-6ex]}\hspace{1em}}  & \multicolumn{1}{c}{\hspace{-1em}\makecell[c]{Model\\[-6ex]}\hspace{1em}} & \multicolumn{2}{c}{\makecell[c]{Results}} \\  
\cmidrule(lr){3-4}  
& & \makecell[c]{SHARP\\parameters} & \makecell[c]{HED\\parameters} \\  
\midrule 
BSS    & \begin{tabular}[c]{@{}c@{}}Transformer\\ BiLSTM-Attention\\ BiLSTM\\ LSTM-Attention\\ LSTM\\ NN\end{tabular} & \begin{tabular}[c]{@{}c@{}}\textbf{0.304 ± 0.092}\\ 0.161 ± 0.089\\ 0.152 ± 0.124\\ 0.120 ± 0.069\\ 0.142 ± 0.133\\ 0.080 ± 0.078\end{tabular} & \begin{tabular}[c]{@{}c@{}}\textbf{0.522 ± 0.060}\\ 0.446 ± 0.086\\ 0.442 ± 0.090\\ 0.385 ± 0.126\\ 0.400 ± 0.115\\ 0.282 ± 0.075\end{tabular} \\\midrule 
BS     & \begin{tabular}[c]{@{}c@{}}Transformer\\ BiLSTM-Attention\\ BiLSTM\\ LSTM-Attention\\ LSTM\\ NN\end{tabular} & \begin{tabular}[c]{@{}c@{}}\textbf{0.154 ± 0.020}\\ 0.185 ± 0.020\\ 0.187 ± 0.028\\ 0.194 ± 0.016\\ 0.189 ± 0.030\\ 0.203 ± 0.018\end{tabular} & \begin{tabular}[c]{@{}c@{}}\textbf{0.105 ± 0.014}\\ 0.122 ± 0.019\\ 0.123 ± 0.020\\ 0.136 ± 0.028\\ 0.132 ± 0.025\\ 0.158 ± 0.016\end{tabular} \\ \bottomrule
\end{tabular}
\end{table}

\section{Conclusions and Discussions}
In this paper, we first establish a database comprising two datasets: the SHARP dataset and the HED dataset. The ARs contained in both datasets are identical. Furthermore, the start and end times for the same AR are identical in both datasets. Subsequently, we adopt an AR-based CV method to divide the SHARP dataset and the HED dataset into training, validation, and testing datasets, while maintaining consistency in the AR distribution between the datasets. We develop six models for forecasting solar flares within 24 hours, using both SHARP parameters and HED parameters. We subsequently perform a comparative analysis of these models, evaluating both their categorical and probabilistic forecasting performance. Ultimately, we select the Transformer model, which exhibits superior performance for both SHARP and HED parameters, to conduct a parameter importance analysis. By employing RFE to prune parameters, we aim to uncover commonly significant parameters that influence categorical forecasting performance (TSS) and probabilistic forecasting performance (BSS) for SHARP and HED parameters, respectively. 
\par The main results of this paper are summarized as follows: (1) In categorical prediction, among the six solar flare forecasting models, the Transformer model performs the best. The Transformer model achieves a TSS of 0.559 ± 0.087 using SHARP parameters, and a TSS of 0.721 ± 0.084 using HED parameters. Additionally, all solar flare forecasting models using HED parameters outperform those using SHARP parameters in terms of categorical forecasting performance. (2) In probabilistic prediction, the Transformer model stands out as the best among the six solar flare forecasting models. It achieves a BSS of 0.304 ± 0.092 when using SHARP parameters and a noticeably higher BSS of 0.522 ± 0.060 with HED parameters. Notably, solar flare forecasting models that employ HED parameters outperform those that utilize SHARP parameters in probabilistic forecasting performance. (3) In the analysis of parameter importance utilizing SHARP parameters, R\_VALUE, SAVNCPP, and ABSNJZH emerge as the top three parameters in both TSS and BSS scores, while AREA\_ACR consistently lags. Notably, R\_VALUE performs the best in both categorical and probabilistic forecasting, highlighting its crucial role in predicting $\ge $ M class flares. (4) In the analysis of parameter importance based on HED parameters, $\mathrm{E_{free}}$ and $\mathrm{H_{C}}$ stand out as the top two in both TSS and BSS scores, with $\mathrm{E_{free}}$ topping the list. This indicates that $\mathrm{E_{free}}$ possesses a strong ability to differentiate between C-class and M/X-class flares.
\par In our study, the R\_VALUE stands out as the best performer among SHARP parameters due to its close correlation with solar flares. The large flares frequently occur near strong and highly sheared PILs \citep{sadykov2017relationships,vasantharaju2018statistical,dhakal2023causes}, and the R\_VALUE focuses on the regions near the PIL. SAVNCPP and ABSNJZH  perform better than TOTUSJH and TOTUSJZ, which implies that the net current and current helicity react sharper to the solar flares than the total quantities. Previous studies have shown that flare/CME-productive ARs are more likely to exhibit non-neutralized currents than flare/CME-quiet ARs \citep{avallone2020electric,liu2024nonneutralized}. Our result suggests that the net current/current helicity in ARs is more closely related to their eruptive potential than the total current/current helicity \citep{dalmasse2015origin,wang2023investigating}, which probably implies the presence of a magnetic flux rope with net currents before the flare. Among the HED parameters, $\mathrm{E_{free}}$ exhibits the optimal performance. In previous studies, the free magnetic energy stored in an AR is a relatively good parameter that is closely related to large flares \citep{emslie2012global,liokati2022magnetic,xu2022evolution}.  The HED regions almost coincide with the PILs with a steep horizontal magnetic gradient (see Figure 1 in \cite{li2024survey}).  Several previous studies have shown that the PIL characteristics are more effective than global characteristics of ARs \citep{sadykov2017relationships,wang2019parameters,sun2021improved}. Moreover, the $\mathrm{E_{free}}$ and $\mathrm{H_{C}}$ parameters within HED region perform better than TOTPOT and TOTUSJH calculated within the entire active region. This suggests that the HED parameters that do not scale with the active region size (such as average quantities or quantities relevant to a specific source region) present a better ability to distinguish between C-class and M/X-class flaring ARs than the SHARP parameters that scale with the size of the active region. In total, our results show that the forecasting based on HED parameters demonstrates higher scores than SHARP parameters, which are consistent with previous studies. 
\par Solar flares significantly affect the production and daily life on Earth. Therefore, accurately predicting their timing and intensity is crucial for the planet's operations. In this paper, we explore the impact of magnetic field characteristics extracted from various regions of the same AR and assess how different models influence the performance of solar flare predictions. Our findings will contribute to the improvement of future research regarding forecasting solar flares. In our future research, we aim to incorporate a broader range of photospheric magnetic field parameters into solar flare predictions. We plan to utilize the data provided by the European Space Agency's (ESA; \citealp{akioka2005l5,lu2019ena}) Vigil L5 mission, as well as Chinese telescopes like the Advanced Space-based Solar Observatory (ASO-S; \citealp{gan2023advanced}) and the Chinese H Solar Explorer (CHASE; \citealp{li2019chinese,cheng2022introduction}) satellite. Our ultimate objective is to enhance the accuracy in forecasting the occurrence time and intensity of solar flares. Furthermore, we are dedicated to developing a real-time prediction platform that leverages the latest observational data and advanced deep-learning algorithms to achieve rapid and accurate predictions of solar flares, contributing to global space weather forecasting.

\section*{Acknowledgments}
We express our gratitude to the anonymous reviewers whose insights and feedback have greatly enhanced the quality of this paper. Our acknowledgments also extend to the dedicated members of the SDO/HMI team for their contributions to the SDO mission.
This work is supported by the B-type Strategic Priority Program
of the Chinese Academy of Sciences (Grant No. XDB0560000), the National Natural Science Foundation of China (Grants No. 12222306, No. 12473056,  No. 11703009, and No. 11803010), the Natural Science Foundation of Jiangsu Province, China (Grants No.
BK20241830, No. BK20170566, and No. BK20201199), and is supported by the Qing Lan Project.

\section*{Data availability} 
The data underlying this paper are available via \dataset[ https://doi.org/10.5281/zenodo.13734431]{The data for Prediction of Large Solar Flares Based on SHARP and HED Magnetic Field Parameters}.

%

\vspace{5mm}


\bibliography{sample631}{}
\bibliographystyle{aasjournal}



\end{document}